\documentclass[12pt,preprint]{aastex}

\newcommand{\exhydra}{EX~Hydrae}
\newcommand{\exhya}{EX~Hya}
\newcommand{\euve}{{\it EUVE\/}}
\newcommand{\pht}{\phantom{2}}
\newcommand{\phispin}{\ifmmode{\phi_{67}}\else{$\phi_{67}$}\fi}
\newcommand{\phiorbit}{\ifmmode{\phi_{98}}\else{$\phi_{98}$}\fi}
\newcommand{\phiorbitmin}{\ifmmode{\phi_{98,{\rm 
min}}}\else{$\phi_{98,{\rm min}}$}\fi}
\newcommand{\phispinfold}{\ifmmode{\phi_{67,{\rm 
fold}}}\else{$\phi_{67,{\rm fold}}$}\fi}
\newcommand{\phiorbitfold}{\ifmmode{\phi_{98,{\rm 
fold}}}\else{$\phi_{98,{\rm fold}}$}\fi}
\newcommand{\kms}{\ifmmode{{\rm km~s}^{-1}}\else{km~s$^{-1}$}\fi}

\newcommand{\lax}{{\lower0.75ex\hbox{ $<$ }\atop\raise0.5ex\hbox{ $\sim$ }}}
\newcommand{\gax}{{\lower0.75ex\hbox{ $>$ }\atop\raise0.5ex\hbox{ $\sim$ }}}

\shorttitle{Accretion Disk Structure of EX Hya}
\shortauthors{Hoogerwerf et al.}

\begin{document}


\title{X-RAY LIGHT CURVES AND ACCRETION DISK STRUCTURE OF EX HYDRAE}


\author{R.\ Hoogerwerf, N.~S.\ Brickhouse}
\affil{Smithsonian Astrophysical Observatory, Harvard-Smithsonian
Center for Astrophysics, 60 Garden Street, MS 31, Cambridge, MA 02138}
\email{rhoogerwerf@cfa.harvard.edu,nbrickhouse@cfa.harvard.edu}

\and

\author{C.~W.\ Mauche}
\affil{Lawrence Livermore National Laboratory, L-473, 7000 East Avenue,
Livermore, CA 94550}
\email{mauche@cygnus.llnl.gov}


\begin{abstract}

We present X-ray light curves for the cataclysmic variable \exhydra\
obtained with the {\it Chandra\/} High Energy Transmission Grating
Spectrometer and the {\it Extreme Ultraviolet Explorer\/} Deep Survey
photometer. We confirm earlier results on the shape and amplitude of
the binary light curve and discuss a new feature: the phase of the
minimum in the binary light curve, associated with absorption by the
bulge on the accretion disk, increases with wavelength.
We discuss several scenarios that could account for this trend and
conclude that, most likely, the ionization state of the bulge gas is
not constant, but rather decreases with binary phase. We also conclude
that photoionization of the bulge by radiation originating from the
white dwarf is not the main source of ionization, but
that it is heated by shocks originating from the interaction between
the inflowing material from the companion and the accretion disk. The
findings in this paper provide a strong test for accretion disk models
in close binary systems.

\end{abstract}




\section{Introduction}

Cataclysmic variables (CVs) are semi-detached, interacting binaries
consisting of a white dwarf and a late-type star (hereafter, the
companion) that overflows its Roche Lobe. In most cases the overflowing
material settles into a disk around the white dwarf before accreting
onto its surface. At the point where the stream of material from the
companion collides/interacts with the accretion disk a ``hot-spot'' or
``bulge'' forms. Much of our understanding of the three-dimensional
structure of CV disks has been obtained from studying their light
curves; for CVs seen nearly edge-on, light curves show broad modulations
(either in emission or absorption, depending on the wavelength regime)
and deep eclipses. The broad modulation is due to the hot-spot orbiting
with the binary system and the eclipses are due to occultation of the
white dwarf and/or hot-spot by the companion \citep[see
e.g.,][]{woo1986}.

A subset of the CVs, the so-called magnetic CVs, contain white dwarfs
with magnetic fields that are sufficiently strong to force the accreting
material away from the orbital plane of the binary and onto one or
both of the white dwarf's magnetic poles. The accreting material
approaches the white dwarf supersonically and so passes through a
stand-off shock before cooling and settling onto the surface of the
white dwarf \citep{aiz1973,kyl1982}. The material in the shock is hot
($\sim 10$--100~keV), so magnetic CVs are powerful X-ray sources.
Magnetic CVs are divided into two subclasses: polars and intermediate
polars (IPs). In polars the stars are tidally locked; this ``static''
configuration results in an accretion flow from the inner Lagrangian
point of the binary along the magnetic field lines of the white dwarf
onto its magnetic pole(s). In IPs, where the white dwarf is spinning
faster than the orbital period, matter also accretes onto the white dwarf
along the magnetic field lines, but it first accumulates in an accretion
disk surrounding the white dwarf.

In this paper we discuss the X-ray light curves of the IP \exhydra\
(hereafter, \exhya) obtained with the {\it Chandra X-ray Observatory\/}
({\it Chandra\/}) High Energy Transmission Grating (HETG) spectrometer
and the {\it Extreme Ultraviolet Explorer\/} ({\it EUVE\/}) Deep Survey
(DS) photometer. \exhya\ consists of a $0.49\pm 0.13~{\rm M}_\odot$ white
dwarf with a spin/rotation period of 67 minutes and a
$0.078\pm0.014~{\rm M}_\odot$ companion
\citep*{fuj1997,beu2003,hoo2004}. The binary system has an orbital
period of 98 minutes 
and an inclination $i=77^\circ $.
\exhya\ light curves have been studied extensively
in the optical \citep*[e.g.,][]{vog1980,ste1983}, 
UV \citep*[e.g.,][]{mau1999,bel2003}, 
EUV \citep*[e.g.,][]{hur1997, bel2002},
and X-rays \citep*[e.g.,][]{ros1988,all1998}.

The X-ray white dwarf light curve, i.e., the observed flux as a
function of the white dwarf spin phase (hereafter 
\phispin ) shows a broad sinusoidal modulation that peaks around
$\phispin \approx 1.0$.  Its amplitude increases with increasing
wavelength of the emission. Two origins have been suggested for this
broad modulation. One is the ``accretion curtain'' model
\citep{ros1988}, in which the pre-shock material absorbs emission
created in the post-shock region. Since the pre-shock material is
confined to the magnetic field, its projected column density changes
as the white dwarf rotates, hence the amount of absorption varies
\citep[see figures 7 and 8 of][]{ros1988}. The other is the
occultation model \citep[e.g.,][]{all1998,muk1999}, in which the
\phispin\ modulation is due to occultation of the emission by the limb
of the white dwarf.

The X-ray binary light curve of \exhya, i.e., the observed flux as a
function of binary phase (hereafter \phiorbit ), shows two
main features: a broad modulation, thought to be due to photoelectric
absorption of white dwarf and accretion column emission by the hot
spot or bulge material on the accretion disk, and a sharp partial
eclipse, due to the occultation of the lower accretion pole of the
white dwarf by the companion \citep[see e.g.,][]{ros1988,muk1998}.

This paper focuses on the binary light curve using data obtained with
{\it Chandra\/} and {\it EUVE\/} and is organized as follows: \S~2
describes the data and the construction of the light curves, \S~3 presents
the binary light curves and a model for the accretion disk
bulge, \S~4 discusses the results, and \S~5 presents the conclusions.

\section{Observations and Reduction}

\subsubsubsection{Chandra}

\exhya\ was observed by \dataset[ADS/Sa.CXO#obs/01706]{{\it
Chandra\/}} using the HETG in combination with the Advanced CCD
Imaging Spectrometer in its spectroscopy layout (ACIS-S) on 2000 May
18 for 60 ks. The observation was continuous and covers $\sim 10$
orbital revolutions of the binary system and $\sim 15$ rotations of
the white dwarf. We reduced the data using the {\it Chandra\/} 
Interactive Analysis of Observations (CIAO version 3.0) software
package\footnote{http://cxc.harvard.edu/ciao/}, making only two
departures from the standard reduction: (1) we turned randomization
off to minimize any artificial line broadening (i.e., we set {\it
rand\_pix\_size\/} equal to zero in the {\it tg\_resolve\_events\/}
tool; note that in CIAO version 3.1 and later {\it rand\_pix\_size\/}
= 0 is the default setting), and (2) we applied a solar system
barycentric correction so that the event times are in Barycentric
Dynamical Time instead of spacecraft time.

Light curves were constructed using the CIAO tool {\it dmextract\/}. We
note that since the HETG grating spectra span several CCDs on the ACIS
detector, it is recommended in the {\it dmextract\/}
documentation\footnote{http://cxc.harvard.edu/ciao/why/lightcurve.html}
that light curves be extracted for each CCD individually so that the
correct Good Time Interval (GTI) table for each CCD is used (the GTIs
usually differ from CCD to CCD). The individual CCD light curves can
then be combined to form the correct light curve, for example, a
spectral order light curve.  For the \exhya\ observation, the GTIs for
ACIS CCDs \#4 through \#9 differed by less than 6.6~s, or $\sim 0.01\%$
of the 60~ks integration time, so we chose to ignore this small
difference and extract the light curves for spectral orders spanning
several CCDs in the same extraction (in this case the first GTI table
available to {\it dmextract\/} is used, i.e., the CCD \#7 GTI, which
contains the zeroth order).

We generated light curves for the combined first orders of the High
Energy Grating (HEG) and Medium Energy Grating (MEG) as a function of
binary phase \phiorbit\ for the optical ephemeris of
\citet{hel1992}. Furthermore, we created light curves for the
following wavelength intervals: 1--5 \AA, 5--10 \AA, 10--15 \AA, and
15--20 \AA. In the remainder of this paper we refer to these light
curves as $\pounds(\lambda)$, where $\lambda$ indicates the wavelength
range.

We also generated light curves for the HETG background and conclude
that they are constant with time and account for less than 0.4\% of
the signal in the source light curves.

\subsubsubsection{EUVE}

\exhya\ was observed by \euve\ \citep{bow1992,bow1994} between 2000 May
2 and 2000 June 15.  The data were extracted from the
HEASARC\footnote{http://heasarc.gsfc.nasa.gov} data archive in the
form of 11 FITS-format events files,
\dataset[ADS/Sa.EUVE#ex_hya__0005021734N]
\dataset[ADS/Sa.EUVE#ex_hya__0005060614N]
\dataset[ADS/Sa.EUVE#ex_hya__0005091840N]
\dataset[ADS/Sa.EUVE#ex_hya__0005130713N]
\dataset[ADS/Sa.EUVE#ex_hya__0005170801N]
\dataset[ADS/Sa.EUVE#ex_hya__0005200756N]
\dataset[ADS/Sa.EUVE#ex_hya__0005241334N]
\dataset[ADS/Sa.EUVE#ex_hya__0005282112N]
\dataset[ADS/Sa.EUVE#ex_hya__0006020541N]
\dataset[ADS/Sa.EUVE#ex_hya__0006061415N]
\dataset[ADS/Sa.EUVE#ex_hya__0006102254N]
which were manipulated with custom IDL software developed over many
years of {\it EUVE\/} observations of cataclysmic variables
\citep[see][and references therein]{mau2002}. Source counts were
collected from a 2 arcmin radius circle centered on EX Hya, while
background counts were collected from a surrounding annulus having an
inner radius of 3 arcmin and an outer radius of 5.385 arcmin (such
that the ratio of source to background areas was 1:5). After
discarding numerous short ($\Delta t < 100$~s) data intervals
comprising less than 1\% of the total exposure, we ``manually''
adjusted the start and stop times of the nominal good time intervals
to include only those times when the source was above Earth's limb and
the background and primbsch/deadtime correction was low. This
filtering resulted in 719 good time intervals for a total exposure of
967 ks.  Event times were corrected from spacecraft time to
HJD, HJD was converted to binary phase using the ephemeris of
\citet{hel1992}, phase-folded source $S$, background $B$, and
primbsch-weighted exposure $\Delta t$ light curves were accumulated,
and the background-subtracted light curves and errors were calculated
as $(S-B/5)/\Delta t$ and $\sqrt{(S+B/25)}/\Delta t$, respectively.

\section{Binary Light Curve}

\begin{figure}
\plotone{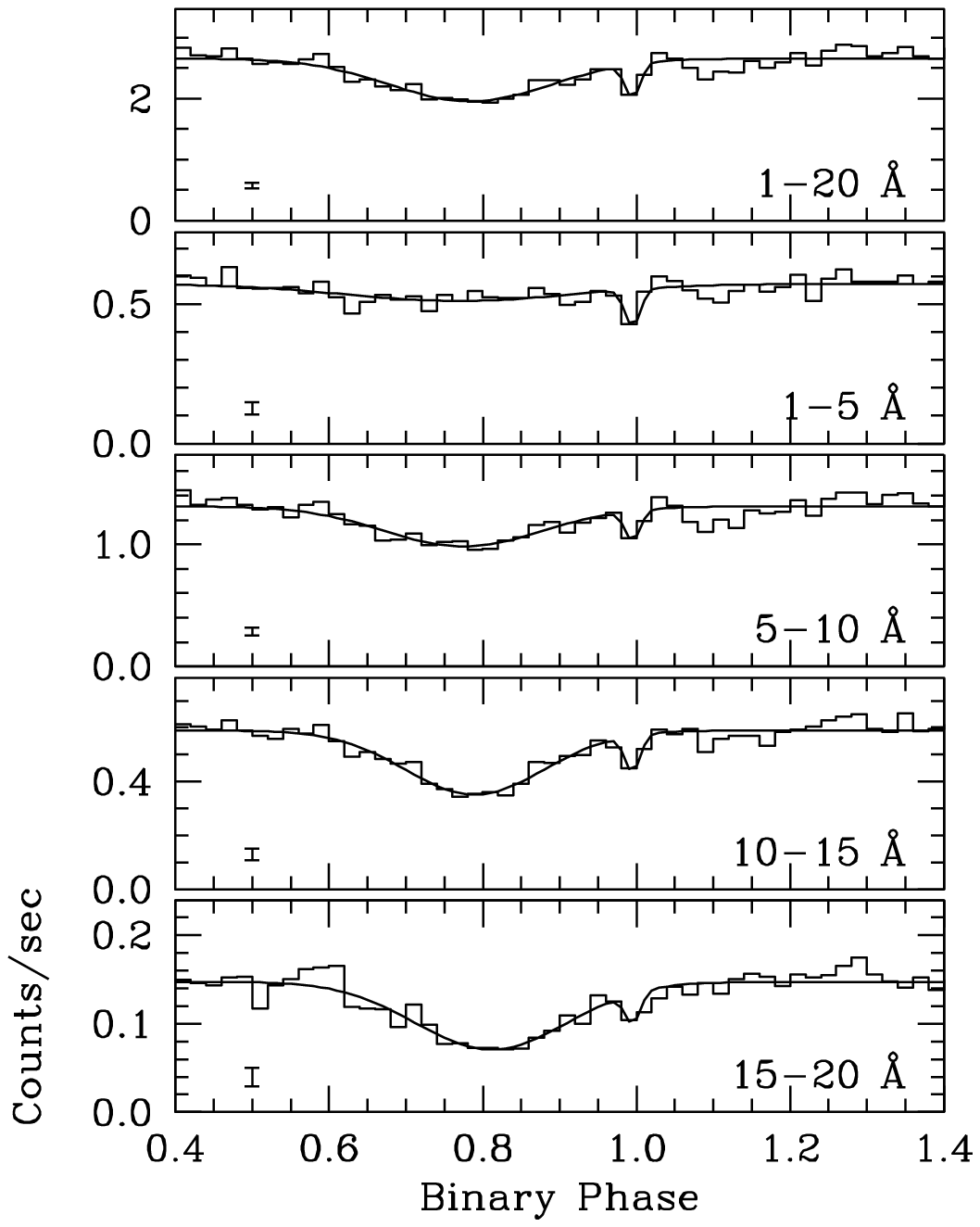}
\caption{EX~Hya binary light curves in the five {\it Chandra\/}
first-order HETG wavelength bands. The average error of each light
curve is indicated in the lower left corner of each panel. Each panel
also shows, as a solid line, the model fit to the data discussed in
\S~3.2, with the fit parameters in Table~\ref{tab:2}. \label{fig:4}}
\end{figure}

Figure \ref{fig:4} shows the {\it Chandra\/} light curves of \exhya\
folded on the binary phase. A narrow partial eclipse is visible around
$\phiorbit = 1.0$, as is the broad modulation centered on $\phiorbit
\sim 0.8$. We discuss the eclipse and the broad modulation separately
in the following two sections.

\subsection{Eclipse}

We measured the position and duration of the eclipse by fitting
$\pounds(\lambda)$ for $\phiorbit \in [0.9,1.1]$, using 10 s bins (or
0.0017 in \phiorbit), with a second-order polynomial to represent the
overall shape of the light curve, and a Gaussian to represent the
eclipse (note that the bins used for the fit are much smaller than
those shown in Fig.~\ref{fig:4}). We found for $\pounds(1$--20~\AA )
that the eclipse is centered on $\phiorbit = 0.9947\pm0.0009$, which
is slightly (31 s) before the time of eclipse predicted by the
ephemeris of \cite{hel1992}, but well within its errors
($\sigma_\phiorbit = \pm0.012$ at the time of the observation). The
eclipse has a FWHM of $(0.024\pm0.002)\times\phiorbit$ or $141\pm
12$~s, and has an eclipse deficit\footnote{The eclipse deficit is
defined by \citet{ros1988} as the total count rate during the eclipse,
defined as 2 times the FWHM, compared to the count rate in two
adjoining regions, each one FWHM wide, on either side of the eclipse.}
of $12.8\%\pm0.4\%$. We find no evidence, at the $2\sigma$ level, for
any dependence of the time of eclipse, its width, or its deficit with
wavelength (see Table~\ref{tab:2}). The FWHM of the eclipse is in good
agreement with the measurement by \citet{muk1998} of $157\pm4$~s based
on 49 eclipses observed with {\it RXTE\/}. Furthermore, the eclipse
deficit obtained from the {\it Chandra\/} data is in rough agreement
with that obtained from {\it EXOSAT\/}, $20\pm4\%$ \citep[for the
1.5--3 keV band;][]{ros1988} and {\it Ginga\/}, $18.2\pm1.7\%$
\citep[for the 1.7--2.8 keV band;][]{ros1991}.

\subsection{Bulge dip}

The broad modulation in the \exhya\ binary light curve is thought to
be due to absorption by the hot spot or bulge on the accretion
disk. In our first attempt to describe the broad modulation, we
modeled the light curves assuming a constant count rate and two
Gaussians: a narrow one centered on $\phiorbit \sim 1$ representing
the eclipse discussed in the previous section, and a broad one
centered on $\phiorbit \sim 0.8$ representing the broad modulation.
Table~\ref{tab:2} lists the centroids, FWHM, and depth of the Gaussian
fit to the broad modulation and the fits are shown in Figure~\ref{fig:4}.

As has been reported before \citep[e.g.,][]{ros1991,all1998}, the
depth of the broad modulation becomes more pronounced for the longer
wavelengths. However, we also find that the phase of the minimum of
the broad modulation changes with wavelength. Both of these trends
continue into the EUV wavelength band, as can be seen from
Figures~\ref{fig:5} and \ref{fig:6}, which show the {\it Chandra \/}
HETG $\pounds(1$--5~\AA), $\pounds(5$--10~\AA), $\pounds(10$--15~\AA),
and $\pounds(15$--20~\AA) light curves and the {\it EUVE\/} DS
$\pounds(70$--180~\AA) light curve. We argue below that the shift in
the phase of the minimum of the broad modulation (hereafter
\phiorbitmin) with wavelength from X-rays to the EUV is most likely
due to a variation with \phiorbit\ in the photoelectric absorption
characteristics of the absorbing material.

\subsection{Modeling the Bulge Dip}

In the following, we assume that the bulge dip can be modeled by a
single column density that fully covers the source of X-ray and EUV
light. In this case, the observed count rate $C(\lambda,\phiorbit)$
can be written as:
\begin{equation}
\label{eq:abs}
C(\lambda,\phiorbit) =
     \bar{F}_0(\lambda)
     \exp \left [ - N_{\rm H}(\phiorbit)
                    \sum\limits_{i=1}^{28} A_{i}
                    \sum\limits_{j=1}^{i} \sigma_{ij}(\lambda)
                                      I_{ij}(\phiorbit)
           \right ],
\end{equation}
where
$\bar{F}_0(\lambda) = F_0(\lambda)\, ARF(\lambda)$,
$F_0(\lambda)$ is the flux (in $\rm photons~cm^{-2}~s^{-1}~\AA^{-1}$)
of the source before it passes through the absorber,
$ARF(\lambda)$ is the effective area of the instrument (in $\rm cm^2$),
$N_{\rm H}(\phiorbit)$ is the H column density (in $\rm cm^{-2}$),
$A_{i}$ is the fractional abundance of element $i$ relative to H, and
$I_{ij}(\phiorbit)$ and $\sigma_{ij}(\lambda)$ are respectively the
ionization fraction and photoelectric cross section of ion $j$ of
element $i$. We assume that the source spectrum is constant with
\phiorbit , that the abundances are solar \citep{gre1998}, and that the
photoelectric cross sections are as given by \citet{ver1995}. With these
assumptions, variations in $C(\lambda,\phiorbit)$ with \phiorbit\ can be
due to changes in the column density and/or the ionization state of the
absorbing material.

To investigate the hypothesis that a changing ionization state could
produce the \exhya\ light curves, we consider two physical processes
that affect the ionization state of the absorbing material: collisional
ionization (CI) and photoionization (PI). For both processes we assume
that the absorber is optically thin and has a homogeneous ionization
balance (but see \S~3.4). With these assumptions, the ionization balance
is given for CI by the temperature $T$ and for PI by the ionization
parameter $\xi = L/nR^2$, where $L$ is the luminosity (in $\rm
erg~s^{-1}$), $n$ is the H density (in $\rm cm^{-3}$), and $R$ is the
separation between the absorber and the source of the ionizing radiation
(in cm). In the CI case, we assume that the ionization fractions are as
given by \citet{maz1998} as a function of $T(\phiorbit)$. In the PI case,
we calculated the ionization fractions with the CLOUDY photoionization
code \citep{fer1996} as a function of $\xi(\phiorbit)$.

In addition to $L$, $n$, and $R$, it is necessary in CLOUDY to specify
the shape of the ionizing spectrum. For this, we used a combination of
(1) a multi-temperature APEC \citep{smi2001} thermal plasma model fit to
the {\it Chandra\/} data, representing the emission from the  accretion
columns, (2) a blackbody spectrum to represent the white dwarf surface
\citep[radius $R=10^9$~cm and temperature $T=10000$~K,][]{eis2002},
and (3) a blackbody spectrum to represent the portion of the white
dwarf surface that is heated by emission from the accretion columns
\citep[effective radius $R=6\times10^8$~cm and temperature
$T=25000$~K,][]{eis2002}. We thus assume that the absorber is
photoionized from only one side by the white dwarf and accretion
columns. The ionizing radiation from the disk and the companion are
negligible and can be ignored.

To determine the spectral shape in the {\it Chandra\/} wavelength
band, we fitted the APEC model to the part of the observation least
affected by absorption, i.e., $\phispin \in [0.9,1.3]$ and $\phiorbit
\in [1.0,1.6]$.  The APEC model consists of 35 temperature components
ranging from $1.2\times10^5$ K to $2.3\times10^8$ K with an emission
measure (EM $= \int n_{\rm e} n_{\rm H} {\rm d}V$) distribution $EM(T)
\propto T^{1/2}$ for $T>10^7$~K and $EM(T) \propto T^{3/2}$ for
$T<10^7$~K, i.e., the typical emission measure distribution for a
plasma cooling mainly through thermal bremsstrahlung for $T > 10^7$~K
and mainly through line emission below $10^7$~K. To make sure that the
model correctly reproduces the ultraviolet and far ultraviolet flux
observed in \exhya , we added an additional temperature component at
$T=3.1\times10^5$~K with an emission measure of
$2\times10^{52}$~cm$^{-3}$ to produce the \ion{O}{6} lines measured by
{\it ORFEUS II\/} \citep{mau1999}.

Consistent with \citet{eis2002}, we set the total luminosity $L=
10^{32}~\rm erg~s^{-1}$ between 1~Ryd and 1000~Ryd, which results in
luminosities of 13, 3, 1.5, and $1.9\times 10^{-11}~\rm erg~cm^{-2}$
$\rm s^{-1}$ in the $\ge 1.0$~keV, 0.28--1.0~keV, 0.067--0.028~keV,
and 0.0136--0.067~keV bands, respectively.  The ionization parameter
$\log \xi$ was sampled from $-4$ to 4 in steps of $\Delta\log\xi =
0.25$, $R$ was fixed at $10^{10}$~cm (see \S~3.4), and the density $n$
varied accordingly. The choices of $R$ and $n$ are arbitrary since it
is only $\xi$ that detemines the ionization fraction.  Since we are
interested only in the optically thin case, we adopted the ionization
balance of the first 'zone/shell' in the CLOUDY calculations, i.e, the
inner edge of the absorber.

\subsection{Fit Results}

To make the fitting process simpler we (1) normalize the light curves
to one for $\phiorbit \in [0.2,0.4]$, where we find no appreciable
absorption, (2) use bins $0.02\times\phiorbit$ in size, (3) increase
the errors on the {\it EUVE\/} light curve by a factor of four, to make
them similar to those of the {\it Chandra\/} light curves (so that they
do not dominate the fitting process), and (4) removed the white dwarf
eclipse from the light curves, i.e., we do not fit between $\phiorbit
\in [0.98,1.04]$.

\begin{figure}
\plotone{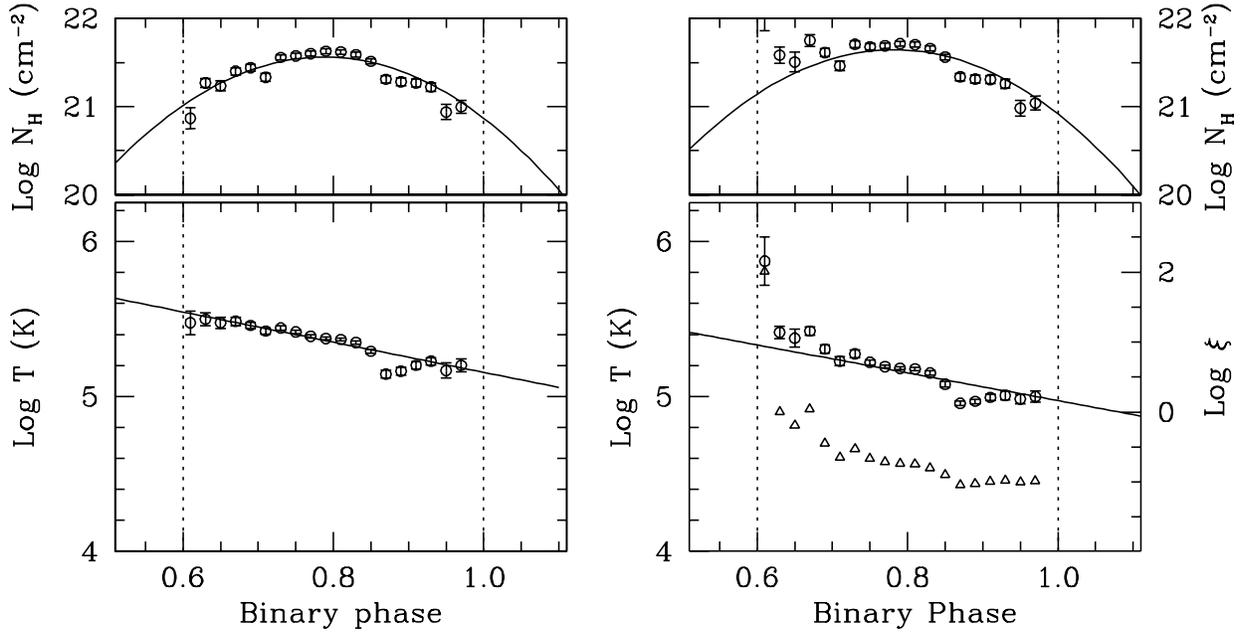}
\caption{
{\it Left (right) panels\/} The best fitting $N_{\rm H}$
and $T$ ($\xi$) for the collisional (photo) ionization model. The
vertical dotted lines in the lower panels indicate the region for
which the absorption is appreciable and can be fitted reliably. The
open circles denote the temperature (ionization parameter) for the
fit. The open triangles denoted the temperature of the gas due to
heating by the PI process. The solid curves show the best fit
temperature (ionization parameter) model (see \S~3.4).
\label{fig:5}}
\end{figure}

Figure~\ref{fig:5} shows the result of fitting the amount of absorption
(Eq.~\ref{eq:abs}) to the {\it Chandra\/} and {\it EUVE\/} light curves
simultaneously, for each \phiorbit\ bin, using as  free parameters the
column density $N_{\rm H}$ and the temperature $T$ for the CI case and the
ionization parameter $\xi$ for the PI case. The figure clearly shows
that, for the region where the absorption is appreciable, i.e.,
$\phiorbit \in [0.6,1.0]$, (1) the variation of the column density can
be described roughly by a Gaussian in \phiorbit\ and (2) $T$ and $\xi$
decline roughly linearly with \phiorbit\ (possibly hitting a floor
around $\phiorbit = 0.85$, after which they remain fairly constant).
The fits are poorly determined outside the $\phiorbit \in [0.6,1.0]$
interval due to the fact that no appreciable absorption is present, so
we do not show the fit parameters in this region.

Based on the trends shown in
Figure~\ref{fig:5} we fitted $T$ and $\xi$ by a linear function:
\begin{equation}
\log X(\phiorbit)  =  c_0 + c_1 \phiorbit,  \label{eq:modela}
\end{equation}
where $X$ represents either $T$ or $\xi$, and fitted $N_{\rm H}$ by a
Gaussian function:
\begin{equation}
N_{\rm H}(\phiorbit) = N_{{\rm H},0}
\exp\left [-(\phiorbit-\phi_{N_{\rm H}})^2/2\sigma_{N_{\rm H}}^2\right ]
\label{eq:nh}
\end{equation}
Table~\ref{tab:3} summarizes the results of the fits, which are shown in
Figure~\ref{fig:6}. While these simple models do not fit the light
curves perfectly, they reproduce, with a minimum number of degrees of
freedom, the essential features of the light curves, including the
change of \phiorbitmin\ with wavelength.

Note that we find a maximum column density $N_{{\rm H},0} \sim
4\times10^{21}~\rm cm^{-2}$, which is a factor four higher than most values
obtained in previous studies \citep[e.g.,][]{hur1997}. This difference
is readily explained by the fact that the other studies assumed neutral
absorbers, while our model of the {\it Chandra\/} and {\it EUVE\/}
light curves requires a partially ionized absorber. Specifically, we
require an absorber with no \ion{H}{1} or \ion{He}{1} opacity; in our
model, most of the absorption in the {\it EUVE\/} wavelength band is 
due to \ion{He}{2}.

Given the parameters inferred for our fits to $N_{\rm H}(\phiorbit)$,
it is possible to draw a schematic picture of the location and extent of
the absorbing material in \exhya . In the sketch shown in Figure~4, we
have assumed that the inner edge of the disk lies at $R_{\rm in}\sim
0.7\times10^{10}$~cm \citep{all1998} and that the outer edge of the disk
lies at $R_{\rm out}\sim 1.5\times10^{10}$~cm \citep[75\% of the white dwarf
Roche lobe radius of $2\times10^{10}$~cm,][]{beu2003,hoo2004}.
The absorber can be drawn on this figure if we assume that its density
$n = N_{\rm H}(\phiorbit)/D(\phiorbit)$ is constant with \phiorbit\ and
its thickness $D(\phiorbit)=R_2-R_1$, where $R_1\ge R_{\rm in}$
and $R_2\le R_{\rm disk}$. Two representative absorbers are shown in
the figure for $R_1=R_{\rm in}$ and $n=5\times 10^{11}~\rm cm^{-3}$
and $R_1=0.75\, R_{\rm disk}$ and $n=1\times 10^{12}~\rm cm^{-3}$.

In the PI case, additional constraints are imposed on the absorber by
the definition of the ionization parameter. Eliminating $n$ between
the relationships $\xi = L/nR^2\sim 3$ and $N_{\rm H}=nD\sim 3\times
10^{21}~\rm cm^{-2}$ gives $D_{10}\sim 0.9\, R_{10}^2$, where $D$ and
$R$ are expressed in units of $10^{10}$~cm. This is further constrained
by the requirement $D/R\ll 1$ so that the ionization parameter does
not vary significantly from the front to the back of the absorber. These
constraints are satisfied only if the absorber is very near the white
dwarf: for $R\le 0.3\, R_{\rm in}$, $D/R\lax 0.3$ and 
$n\gax 4\times 10^{12}~\rm cm^{-3}$. However, this analysis requires that
the PI absorber be optically thin to the ionizing radiation, while
Figure 3 shows that during the dip the {\it EUVE\/} light curve is
extinguished by approximately 95\%, corresponding to an optical depth
$\tau \approx 3$.

\subsection{Optically Thick PI Model}

To allow the possibility of an optically thick PI absorber, we ran a set
of full CLOUDY models to account for the dilution of the radiation field
and the resulting change of the ionization balance with depth into the
absorber. The dashed curve in the middle panel of 
Figure~\ref{fig:6} shows the full CLOUDY light curves based on
our best fit paramaters of the optically thin model for $R = 10^{10}$~cm.
This clearly shows that including the optical depth effects and physical
size of the absorber results in more absorption, particularly at the
longer wavelengths. We are able to produce full CLOUDY models that fit
our light curves as well as we did for the CI and optically thin PI
models, but these models require that the inner radius of the
absorber be at $R\sim 0.4\times 10^{10}$~cm, i.e., smaller than the
inner radius of the accretion disk. While the accretion curtain is a
source of absorption at this close distance to the white dwarf, this
absorption should be visible only on the white dwarf period and not on
the binary period. We thus conclude that the white dwarf in \exhya\ is
not luminous enough to ionize the absorber by photoionization processes
alone.

\subsection{Heating of the Absorber}

We can solve the problems with the PI model described above by imposing
a base temperature to the absorber of order $10^5$~K. This ensures
that no \ion{H}{1} and \ion{He}{1} and almost no \ion{He}{2} is present
in the absorber, as is required by the ionization balances found by the
CI and PI models. This is crucial, since even small traces of these
ions result in large amounts of absorption in the {\it EUVE\/}
waveband. Since this base temperature is significantly higher than the
temperature of the outer disk \citep[3000--7000~K,][]{fra1981,eis2002},
we conclude that viscous heating alone is insufficient to keep the
absorber hot. Other sources of heat, such as shocks due to the
interaction between the accretion stream and the accretion disk, are
needed.

\section{Discussion}

As shown above, the dominant source of ionizing radiation in \exhya ,
the white dwarf and accretion columns, are not sufficiently luminous to
ionize the accretion disk bulge by photoionization alone. Therefore, the
bulge must be ionized through a different process and, assuming that it
is in, or close to, CI equilibrium, its temperature is of order $\log
T({\rm K}) \sim 5.3$ or $T \sim 200,000$~K. Furthermore, to reproduce
the shift of \phiorbitmin\ with wavelength, the bulge must have a
temperature gradient that steadily decreases from $\log T({\rm K})
\sim 5.5$ at $\phiorbit \sim 0.6$ to $\log T({\rm K}) \sim 5.2$ at
$\phiorbit \sim 0.9$. We speculate that this temperature gradient is
caused by the interaction, e.g., shocks, between the stream of
accreting material from the companion and the accretion disk.

Our simple analysis in \S~3.4 places the location of this interaction
region on the disk at $\phiorbit\approx$ 0.6--1.0 corresponding to a
range of angles of approximately $0^\circ$--$130^\circ $ from the
line joining the two stars. The location and shape of the absorber is
uncertain because we measure only the column density along the line of
sight, although Figure~4 shows two possibilities that are consistent
with our data. In interpreting this figure, it must be kept in mind 
that what we are seeing in the X-ray and EUV wavelength bands is only that 
portion of the accretion bulge that rises an angle of $13^\circ$ 
above the orbital plane.

Hydrodynamical simulations of close binary systems show that the
inflowing material from the companion star forms a spiral-like
structure in the accretion disk instead of the canonical hot spot on
the edge of the disk \citep*[e.g.,][]{bis1998, mak2000, sat2003}. The
location of such spiral structures appears to coincide with the
position of the \exhya\ absorber. Furthermore, simulations show that
shocks occur in regions where the inflowing material interacts with
the accretion disk. These shocks might be responsible for the heating
needed in our \exhya\ absorber model. Unfortunately, most of the
hydrodynamic simulations do not include the effects of heating and
cooling on the accretion disk structure \citep[but
see,][]{sat2003}. The \exhya\ results presented in this paper provide
an excellent constraint for new hydrodynamic simulations of close
binary systems.

\subsubsubsection{Visibility of Absorber}
A hot absorber like the one in \exhya\ will generate its own emission,
mainly in the UV band, the strength of which is given by the emission
measure. Unfortunately, we know only the column density and
temperature along a slice through the absorber. It is unknown what the
total volume of the absorber is and whether the temperature
distribution is similar throughout the absorber. Furthermore, the
emission measure depends on the density of the absorber, which is also
an unknown. Estimates and assumptions for all these quantities can be
made, but the result will be highly unreliable. To our knowledge there
is no mention in the literature that the absorber has ever been
observed in emission.

We do however, see evidence of the absorber in {\it ORFEUS II\/} data
\citep{mau1999} and, more clearly, in {\it FUSE\/} data, for which the
analysis is in progress, as line absorption features of the \ion{O}{6}
$\lambda\lambda1032,1038$ doublet lines and the \ion{C}{3}
$\lambda977$, $\lambda1175$ lines at binary phases $0.6 < \phiorbit <
1.0$. The absorption features are not visible at other binary
phases. This range of binary phases agrees well with the absorber
described in this paper. Furthermore, the fact the \ion{O}{6} shows
line absorption confirms that the absorber contains highly ionized
oxygen.

\begin{figure}
\plotone{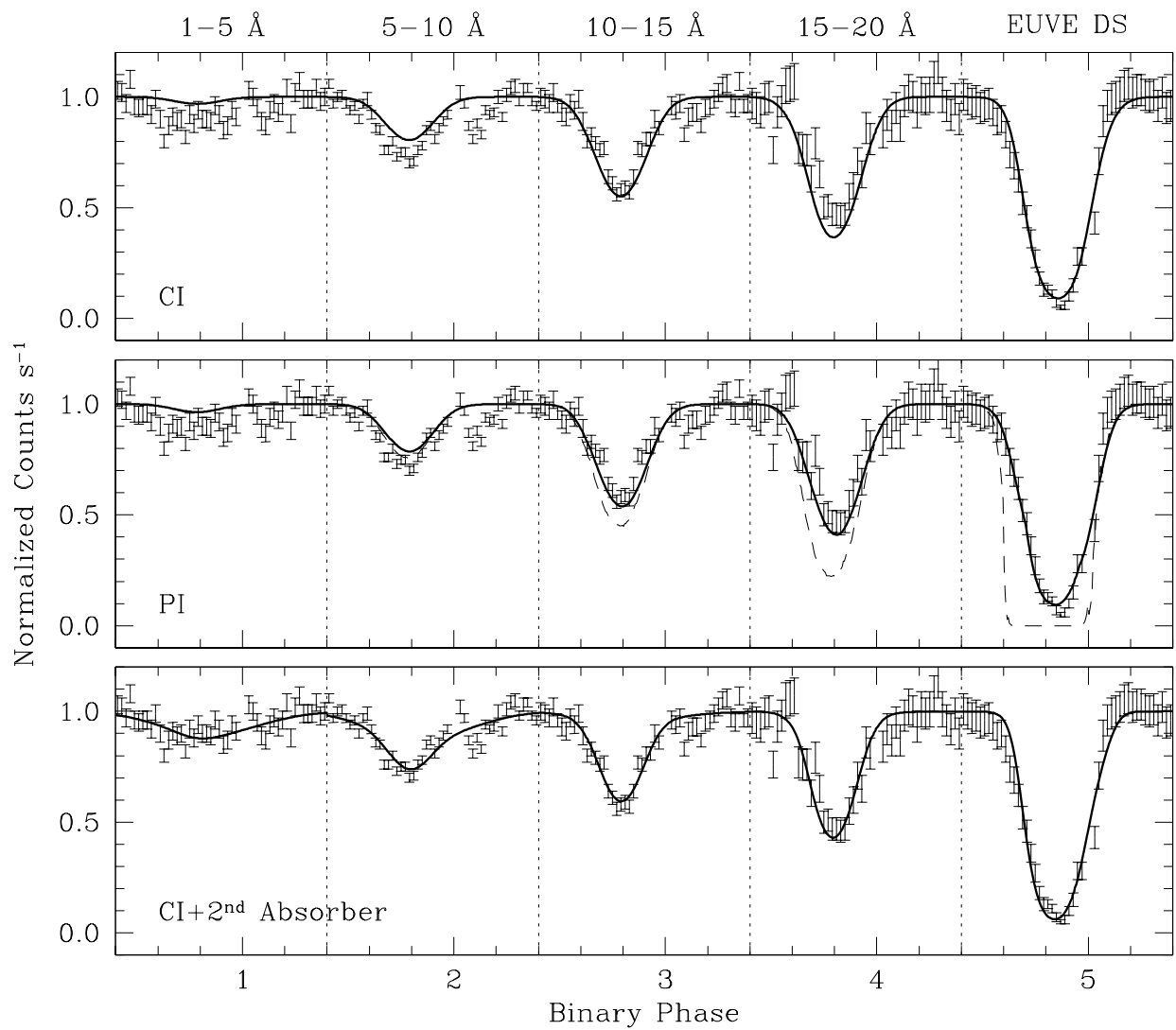}
\caption{
EX~Hya binary light curves in the four {\it Chandra\/} HETG and {\it
EUVE\/} DS wavelength bands. The dotted vertical lines separate the
different wavelength bands. {\it Top panel\/} shows the best fitting
CI model and {\it middle panel\/} shows the best fitting PI model. The
dashed line in the {\it middle panel\/} shows the light curves
predicted by the full CLOUDY calculation discussed in \S~3.5. {\it
Bottom panel\/} shows the light curve produced by the CI model with an
additional absorber.
\label{fig:6}}
\end{figure}

\subsubsubsection{Short Wavelength Absorption}
Figure~\ref{fig:6} shows that both the CI and PI model fits do not
perform very well in the 1--5~\AA\ band and the 5--10~\AA\ band: both
models tend to underpredict the amount of absorption. We found that
adding a second absorber to our model improves the fit: it lowers the
reduced $\chi^2$ from $\sim4$ to $\sim2$ (see bottom panel
Fig.~\ref{fig:6}). The second absorber was modeled with a Gaussian for
the column density and a single temperature or ionization
parameter. The results are that this absorber needs to be highly
ionized [$\log T({\rm K}) \sim 7$ or $\log \xi \sim 2.7$] and have a
large column $\log N_{\rm H}({\rm cm}^{-2}) \sim 23$. Using similar
arguments as in \S~3.4 we can constrain the position of the absorber,
in the PI case, to be very close ($R<4\times10^8$~cm) to the source of
the ionizing emission, i.e., basically at the shock front. It is hard
to understand why this feature would appear on the binary period
rather than the white dwarf period. If real, the second absorber is
most likely to be near the inner edge of the accretion disk, somehow
uncoupled to the magnetic field of the white dwarf. A long observation
of \exhya\ could confirm the existence of a second absorber and
perhaps better constrain its location.

\section{Conclusions}
We report a new feature in the binary light curves of the magnetic
cataclysmic variable \exhya. Based on {\it Chandra}\ HETG and {\it
EUVE}\ DS data we have shown that the phase of the minimum in the
broad modulation, associated with the bulge or hot spot on the
accretion disk, increases with wavelength. Collisional ionization and
photoionization models explain this characteristic as a change in
ionization state of the bulge plasma with binary phase. However, the
ionizing radiation originating on the white dwarf and its accretion
columns is insufficient to account for the ionization state of the
bulge. Moreover, the required ionization state also excludes viscous
heating in the accretion disk as the main source of ionization. We
thus conclude that the bulge plasma is heated by shocks resulting from
the interaction between the inflowing material from the companion and
the accretion disk.



\acknowledgments We thank John Raymond for comments and suggestions
and Yair Krongold and Jonathan Slavin for help setting up with
CLOUDY. We also thank Gary Ferland for a bug fix in and advice on
CLOUDY related to this paper and the referee for a critical reading of
the manuscript. We thank H.~Tananbaum for the generous grant of
Director's Discretionary Time that made possible the {\it Chandra\/}
observations of EX~Hya. This research has made use of data obtained
from the High Energy Astrophysics Science Archive Research Center
(HEASARC), provided by NASA's Goddard Space Flight Center. We
acknowledge support from NASA through Chandra grants GO2--3018X and
GO3-4017X. NB was supported by NASA contract NAS8-39073 to the Chandra
X-ray Center. CWM's contribution to this work was performed under the
auspices of the U.S.~Department of Energy by University of California
Lawrence Livermore National Laboratory under contract
No.~W-7405-Eng-48.

\clearpage


\onecolumn

\begin{figure}
\begin{minipage}{8cm}
\plotone{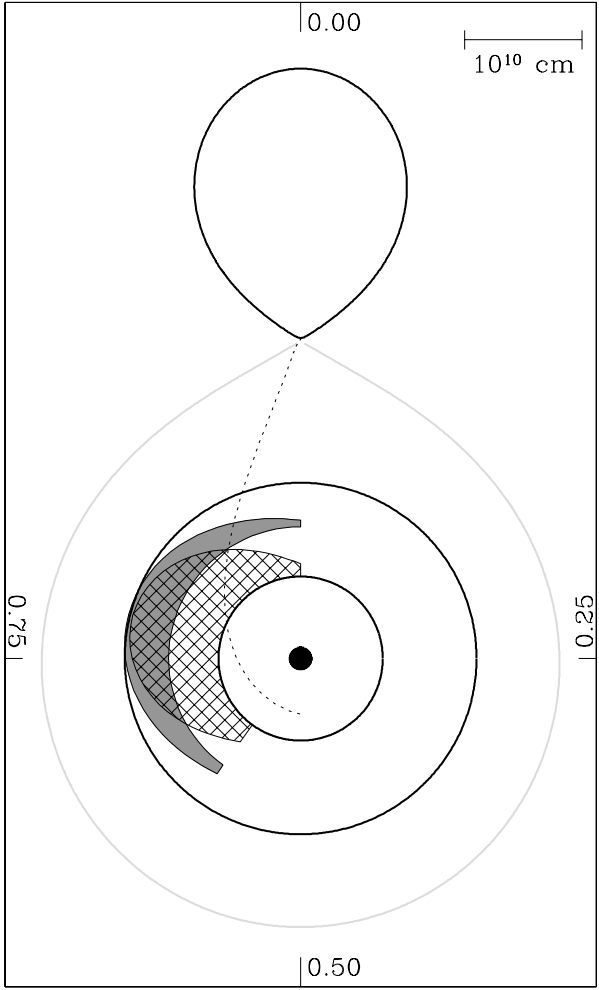}
\end{minipage}
\caption{
Schematic of the \exhya\ binary, showing the companion, the white dwarf
({\it small solid circle\/}), the Roche limit of the white dwarf ({\it
light gray curve\/}), the inner and outer disk radii ({\it circles\/}),
the ballistic stream ({\it dotted curve\/}), and the location of the
absorber for the two cases discussed in \S~3.4. The binary phases are
indicated, as is the scale.
\label{fig:7}}
\end{figure}

%
%
%

\clearpage

\twocolumn

%
%
\begin{deluxetable}{lcccccc}
\tablewidth{0pt}
\tabletypesize{\footnotesize}
\tablecaption{Binary Light Curve Fit Parameters}
\tablehead{
\colhead{}&
\multicolumn{3}{c}{Broad Modulation}&
\multicolumn{3}{c}{Eclipse}\\
\colhead{Wavelength Range}&
\colhead{Centroid}&
\colhead{FWHM}&
\colhead{Depth}&
\colhead{Centroid}&
\colhead{FWHM}&
\colhead{Deficit}\\
\colhead{(\AA)}&
\colhead{(\phiorbit)}&
\colhead{(\phiorbit)}&
\colhead{(\%)}&
\colhead{(\phiorbit)}&
\colhead{(\phiorbit)}&
\colhead{(\%)}
}
\startdata
\phantom{1}1--20           & 0.784$\pm$0.002 & 0.250$\pm$0.006 & 
26.3$\pm$0.5 & 0.995$\pm$0.001 & 0.024$\pm$0.002 & 12.8$\pm$0.4 \\
\phantom{1}1--\phantom{1}5 & 0.769$\pm$0.015 & 0.354$\pm$0.050 & 
10.7$\pm$1.0 & 0.994$\pm$0.001 & 0.020$\pm$0.004 & 17.5$\pm$1.2 \\
\phantom{1}5--10           & 0.775$\pm$0.003 & 0.244$\pm$0.009 & 
25.0$\pm$0.7 & 0.996$\pm$0.001 & 0.024$\pm$0.003 & 11.9$\pm$0.6 \\
10--15                     & 0.790$\pm$0.003 & 0.214$\pm$0.007 & 
40.3$\pm$1.0 & 0.992$\pm$0.002 & 0.025$\pm$0.005 & 12.2$\pm$0.9 \\
15--20                     & 0.811$\pm$0.004 & 0.232$\pm$0.012 & 
52.0$\pm$1.9 & 1.000$\pm$0.002 & 0.017$\pm$0.008 & 14.4$\pm$2.6 \\
\enddata
\label{tab:2}
\end{deluxetable}

%
%
\begin{deluxetable}{lllll}
\tablewidth{0pt}
\tabletypesize{\tiny}
\tablecaption{Collisional and Photoionization Model Parameters}
\tablehead{
\colhead{}&\colhead{}&\colhead{}&\multicolumn{2}{c}{Including second absorber}\\
\colhead{}&\colhead{CI}&\colhead{PI}&\colhead{CI}&\colhead{PI}
}
\startdata
$\log N_{{\rm H},0}$      &    21.59(1)&   21.67(1) &   21.50(1) &   21.55(2)\\
$\phi_{N_{\rm H}}$        &\pht0.786(3)&\pht0.782(3)&\pht0.790(3)&\pht0.764(4)\\
$\sigma_{N_{\rm H}}$      &\pht0.108(3)&\pht0.108(2)&\pht0.094(2)&\pht0.097(2)\\
$c_0$                     &\pht6.13(2) &\pht2.15(3) &\pht6.27(1) &\pht3.34(3)\\
$c_1$                     &\pht0.97(4) &\pht1.99(4) &\pht1.23(2) &\pht3.66(5)\\
$\log \hat{N}_{{\rm H},0}$&            &            &   22.79(8) &   23.1(1)\\
$\phi_{\hat{N}_{\rm H}}$  &            &            &\pht0.85(1) &\pht0.87(1)\\
$\sigma_{\hat{N}_{\rm H}}$&            &            &\pht0.23(2) &\pht0.22(2)\\
$c_3$                     &            &            &\pht7.04(4) &\pht2.69(7)\\
$\chi^2/\nu$              &\pht4.15    &\pht3.79    &\pht2.32    &\pht2.09\\
\enddata
\label{tab:3}
\end{deluxetable}

%

\end{document}